\documentclass[usenatbib]{mn2e}

\usepackage{graphicx}
\usepackage{mathptmx}
\usepackage{subfigure}

\begin{document}

\title[Deep 610-MHz GMRT observations of the Spitzer extragalactic First Look Survey field -- I.]{Deep 610-MHz GMRT observations of the Spitzer extragalactic First Look Survey field -- I.  Observations, data analysis and source catalogue}

\author[T. Garn et al.]{Timothy Garn\thanks{E-mail: tsg25@cam.ac.uk},
                        David A.\ Green,
                        Sally E.\ G.\ Hales,
                        Julia M.\ Riley,
                        Paul Alexander\\
Astrophysics Group, Cavendish Laboratory, 19 J.~J.~Thomson Ave., Cambridge CB3~0HE}

\maketitle

%-----------------------------------------------------------------------------
\begin{abstract}
Observations of the Spitzer extragalactic First Look Survey field taken at 610~MHz with the Giant Metrewave Radio Telescope are presented.  Seven individual pointings were observed, covering a total area of $\sim$4~deg$^{2}$ with a resolution of $5\farcs8 \times 4\farcs7$, PA $60^\circ$.  The r.m.s.\ noise at the centre of the pointings is between 27 and 30~$\mu$Jy before correction for the GMRT primary beam.  The techniques used for data reduction and production of a mosaicked image of the region are described, and the final mosaic, along with a catalogue of 3944 sources detected above $\sim5\sigma$, are presented.  The survey complements existing radio and infrared data available for this region.
\end{abstract}


%------------------------------------------------------------------------------%
\section{Introduction}

The Spitzer First Look Survey was the first major scientific program carried out by the Spitzer Space Telescope \citep{Werner04}, observing for $\sim$110 hours of Director's Discretionary Time in December 2003.  The aim of the extragalactic component of the First Look Survey (xFLS) was to study a region with low Galactic background to a significantly deeper level than any previous large-area infrared survey, in order to accurately characterise the dominant infrared source populations.  Observations were made over four square degrees centred on $17^{\rm h}18^{\rm m}00^{\rm s}$, $+59^\circ30'00''$ (J2000 coordinates, which are used throughout this paper) with two instruments -- the Infrared Array Camera \citep[IRAC,][]{Fazio04}, using the 3.6, 4.5, 5.8 and 8~$\mu$m bands, and the Multiband Imaging Photometer for Spitzer \citep[MIPS,][]{Rieke04} at 24, 70 and 160~$\mu$m.  The data have been processed and maps and source catalogues are currently available (\citealp[IRAC --][]{Lacy05}; \citealp[24~$\mu$m --][]{Fadda06}; \citealp[70 and 160~$\mu$m --][]{Frayer06}).

Complementary observations have been taken at a range of wavelengths to fully exploit the new deep infrared data.  Deep optical surveys of the region have been completed in the $R$-band \citep[KPNO~4~m,][]{Fadda04}, and in the $u^{*}$- and $g$-bands \citep[CFHT~3.6~m,][]{Shim06}.  The xFLS region was covered by the early data release of the Sloan Digital Sky Survey \citep{Stoughton02}.  A further redshift survey targeting selected 24-$\mu$m sources was made with the MMT/Hectospec fiber spectrograph \citep{Papovich06}, and a total of 1587 redshifts are publicly available.

There are two existing 1.4-GHz radio surveys of the xFLS region, made with the Very Large Array (VLA) in B-array configuration, and the Westerbork Synthesis Radio Telescope (WSRT).  The VLA survey covered $\sim$4~deg$^{2}$ with a resolution of $5''$, to a 1$\sigma$ depth of $\sim$23$~\mu$Jy \citep{Condon03} and contains 3565 sources, while the WSRT survey \citep{Morganti04} concentrated on $\sim$1~deg$^{2}$ of the xFLS region to a depth of 8.5~$\mu$Jy~beam$^{-1}$ with a resolution of $14'' \times 11''$, PA $0^\circ$, and contains 1048 sources.

There is a tight correlation between the far-infrared and radio luminosities of galaxies, which has been known for many years \citep[see e.g.][]{Helou85, Condon92}.  The correlation applies to both the local \citep{Hughes06} and global \citep{Murphy06} properties of galaxies in the local and distant universe \citep{Garrett02, Gruppioni03, Chapman05, Luo05}.  There have been two previous studies of the IR/radio correlation in the xFLS region \citep{Appleton04, Frayer06}.  Appleton found no evidence for a variation of the correlation with redshift, while Frayer, using more sensitive infrared data, found a decrease in the infrared/radio flux density ratio with $z$.  To study the luminosity variation it is necessary to accurately $k$-correct the observed radio flux densities to their rest-frame values using the spectral index $\alpha$ (where $\alpha$ is here defined so that flux density scales with frequency as $S \propto \nu^{-\alpha}$).  Both previous works in this region assumed that all sources have the same radio spectral index (taken to be $\alpha$~=~0.7 and 0.8 respectively), but in order to perform this correction accurately it is necessary to have a detection in at least two radio frequencies.

The large amount of existing data on this field makes it a good candidate for a further deep radio survey.  We have imaged the xFLS region at 610~MHz with the Giant Metrewave Radio Telescope (GMRT), reaching an r.m.s.\ noise level of around 30~$\mu$Jy before primary beam correction.  Our survey has a comparable resolution ($5\farcs8 \times 4\farcs7$ compared with $5''$) to the VLA 1.4-GHz survey, allowing a direct comparison between source properties at the two wavelengths to be made.  Our observations are deeper than the VLA survey over most of the image, for a typical radio source with $\alpha~=~0.8$.

\begin{figure}
  \includegraphics[width=8cm]{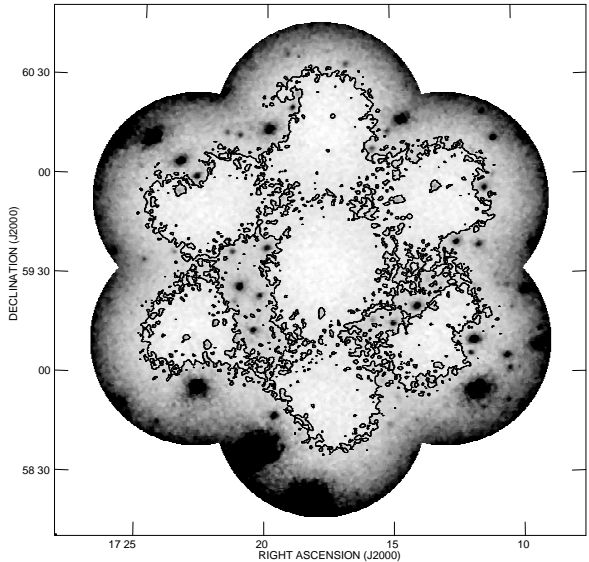}
  \caption{The seven pointings observed with the GMRT.  Pointing centres are spaced by 43 arcmin.  The greyscale represents the r.m.s\ noise level calculated by Source Extractor (see Section~3), after correcting each pointing for the primary beam response of the GMRT and mosaicking the pointings together.  The 45~$\mu$Jy contour (equivalent to the 23~$\mu$Jy noise level of the VLA 1.4~GHz survey, for a source with spectral index of 0.8) has been plotted.}
  \label{fig:sevenpoints}
\end{figure}

\begin{figure}
  \includegraphics[width=8cm]{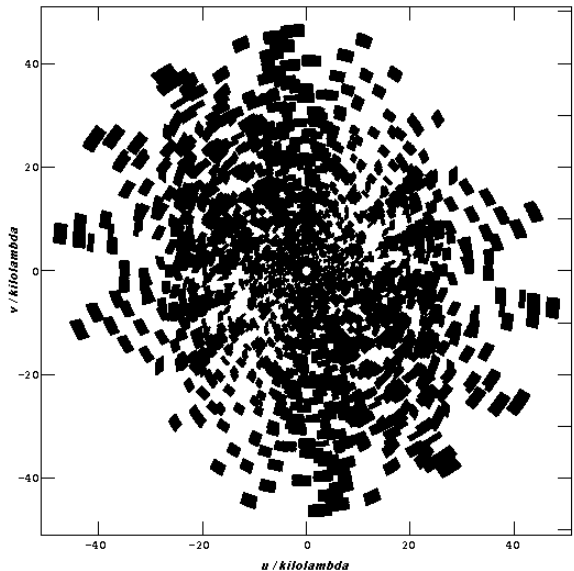}
  \caption{The $uv$ coverage for the central pointing.  Baselines less than 1~k$\lambda$ were not used in imaging and have been omitted.}
  \label{fig:uvcoverage}
\end{figure}

Section~2 describes the observations, and data reduction techniques.  In Section~3 we present the mosaicked image and source catalogue.  We show a selection of extended objects visible in our mosaic, and discuss the artefacts seen near bright sources.  In Section~4 we compare our catalogue with the VLA survey, in order to test the positional accuracy of our data. Further analyses of the data, including detailed comparisions with Spitzer data will be made in forthcoming papers.  Appendix~A presents some technical details of corrections made to the observed GMRT data, and Appendix~B gives details of a correction made to the integrated flux densities of sources.

%------------------------------------------------------------------------------%
\section{Observations and data reduction}

Observations of the xFLS region were made at 610~MHz with the GMRT \citep{Ananthakrishnan05}, located near Pune, India, over four days in March 2004.  Seven pointings were observed, in a hexagonal grid centred on $17^{\rm h}18^{\rm m}00^{\rm s}$, $+59^\circ30'00''$ as shown in Fig.~\ref{fig:sevenpoints}.  The pointings were spaced by 43 arcmin, approximately the half-power bandwidth of the GMRT primary beam at 610~MHz, which gives nearly uniform noise coverage over most of the xFLS region.

Observations of 3C48 or 3C286 were made at the beginning and end of each observing session in order to calibrate the flux density scale.  The AIPS task SETJY was used to calculate flux densities at 610~MHz of 29.4 and 21.1~Jy respectively, using an extension of the \citet{Baars77} scale.  Two sidebands were observed, each of 16~MHz with 128 spectral channels, and a 16.9~s integration time was used.  Each field was observed for $\sim$200~min over four 10~hour observing sessions made between March 23rd and 27th.  The observations consisted of interleaved 20~min scans of each pointing, in order to improve $uv$ coverage.  A nearby phase calibrator, J1634$+$627, was observed for three minutes between the scans of each pointing to monitor any time-dependant phase and amplitude fluctuations of the telescope.  The measured phase of J1634$+$627 varied smoothly between observations, with a typical variation between calibrator observations of less than $40^\circ$\ for the long baseline antennas and below $10^\circ$\ for the short baseline antennas.

Standard AIPS tasks were used to flag bad baselines, antennas, channels that were suffering from narrow band interference, and the first and last 16.9~s integration period of each scan.  A bandpass correction was applied using the flux calibrators, for each antenna.  A pseudo-continuum channel was then made by combining the central ten channels together, and an antenna-based amplitude and phase calibration created using the observations of J1634$+$627.  This calibration was applied back to the original 128 channel data set, which was compressed into 11 channels, each containing ten of the original spectral channels (so the first and last few channels, which tended to be the noisiest, were discarded).  The small width of these new channels ensures that bandwidth smearing is not a problem, and all 11 channels could be individually inspected to remove additional interference.  After the flagging and calibration was complete the two sidebands were combined into a single data set (see Appendix~A1) to improve the $uv$ coverage.  The coverage for the central pointing is shown in Fig.~\ref{fig:uvcoverage}.

\begin{figure*}
  \includegraphics[width=17cm]{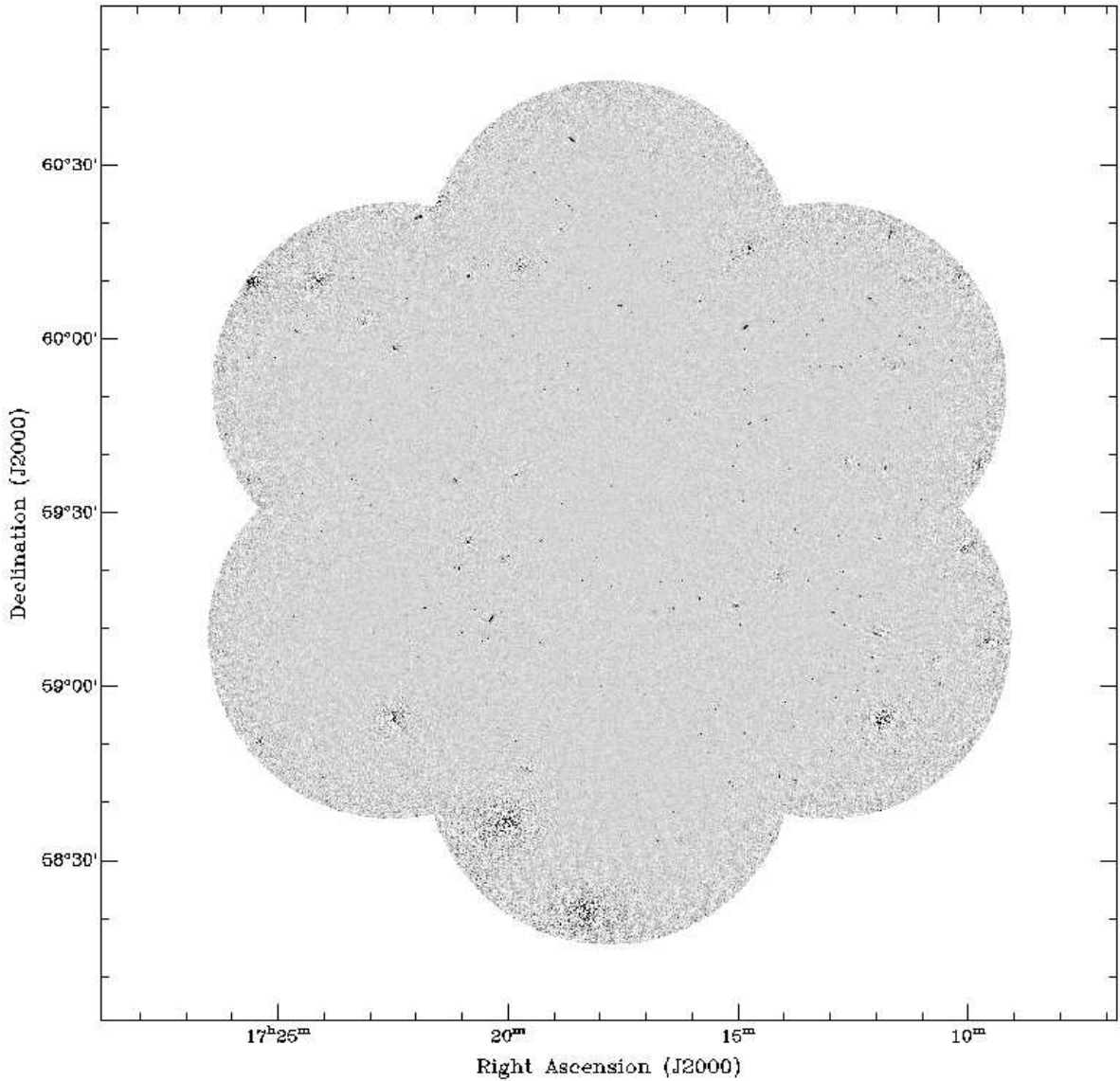}
  \caption{The 610-MHz image of the Spitzer extragalactic First Look Survey field.  The greyscale ranges between $-0.2$ and 1~mJy~beam$^{-1}$.  Artefacts are visible around the brightest few sources.}
  \label{fig:mosaic}
\end{figure*}

The large field of view of the GMRT leads to significant errors if the whole field is imaged directly, due to the non-planar nature of the sky.  To minimise these errors, each pointing was broken down into 19 smaller facets which were imaged separately, with a different assumed phase centre, and then recombined to deal with the transformation from planes to the sphere.  Images were made with an elliptical synthesised beam with size $5\farcs8 \times 4\farcs7$, position angle~$+60^\circ$, with a pixel size of $1\farcs5$ to ensure that the beam was well oversampled.

The GMRT has a large number of small baselines, due to the cluster of 14 antennas in the central 1~km of the array.  This dominates the $uv$ coverage and affects the beam shape, so baselines less than 1~k$\lambda$ were omitted in the imaging.  The images went through three iterations of phase self-calibration at 10, 3 and 1 minute intervals, and then a final round of self-calibration correcting both phase and amplitude errors.  The overall amplitude gain was held constant in order not to alter the flux density of sources.  The initial self-calibration step corrected the phase errors by up to $10^\circ$, with later self-calibration making much smaller changes and having a smaller effect on the r.m.s.\ image noise.

Two problems were identified and corrected during imaging (see Appendix~A for further details).
\begin{enumerate}
  \item An error in the time-stamps of the $uv$ data, and hence the $uvw$ coordinates, was corrected using a custom-made AIPS task.
  \item The GMRT primary beam centre was shifted by $\sim$2.5~arcmin to compensate for a position-dependant flux density error.
\end{enumerate}

The theoretical r.m.s.\ noise of each pointing, before primary beam correction, is
\begin{equation}
  \sigma = \frac{\sqrt{2}T_{\rm{s}}}{G\sqrt{n(n-1)N_{\rm{IF}} \Delta \nu \tau}}
  \label{eq:noise}
\end{equation}
where $T_{\rm s} \approx 92$~K is the system temperature, $G \approx 0.32$~K~Jy$^{-1}$ is the antenna gain -- values taken from the GMRT website\footnote{{\tt {\scriptsize http://www.gmrt.ncra.tifr.res.in/gmrt\_hpage/Users/Help/help.html}}} -- $n$ is the number of working antennas, which was typically 28 during our observations, $N_{\rm IF}=2$ is the number of sidebands, $\Delta\nu$ = 13.75~MHz is the effective bandwidth of each sideband, and $\tau \approx 12000$~s is the average integration time per pointing.

The final images have an r.m.s.\ noise of between 27 and 30~$\mu$Jy~beam$^{-1}$ before primary beam correction, which is very close to the theoretical limit of $\sim$26~$\mu$Jy, although dynamic range issues limit the quality of the images near bright sources where the noise is greater.  The seven pointings were corrected for the primary beam of the GMRT, taking into account the offset beam position as discussed in Appendix~A2.  The beam correction was performed using an 8th-order polynomial, with coefficients taken from \citet{Kantharia01}.  The pointings were then mosaicked together, weighting the final image by the r.m.s.\ noises of each individual pointing, and the mosaic was cut off at the point where the primary beam correction factor dropped to 20\% of its central value.  Figure~\ref{fig:sevenpoints} illustrates the variation in noise across the mosaic.  The noise level is smooth and around 30~$\mu$Jy across the interior of the map, and increases towards the edges to about 150~$\mu$Jy where the primary beam correction was 20\%.

%------------------------------------------------------------------------------%
\section{Results and discussion}

\begin{figure}
  \includegraphics[width=8cm,height=7cm]{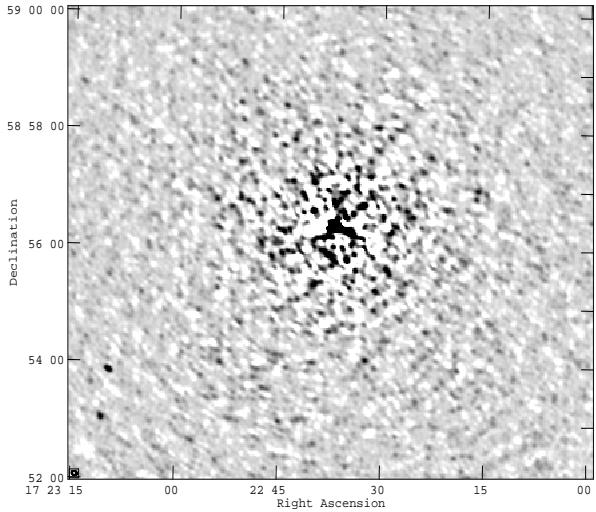}
  \caption{Artefacts visible around a 294 mJy source near $17^{\rm h}22^{\rm m}40^{\rm s}$, $+58^\circ56'00''$.  The greyscale ranges between $-0.2$ and 1~mJy~beam$^{-1}$.}
  \label{fig:brightsource}
\end{figure}

The final mosaic is shown in Fig.~\ref{fig:mosaic}.  There are phase artefacts visible around the brightest sources, which have not been successfully removed during self-calibration.  An enlarged image of one of the bright sources is shown in Fig.~\ref{fig:brightsource}.  It is thought that the artefacts are due to an elevation-dependant error in the position of the GMRT primary beam (see Appendix~A2), which will lead to image distortion near the brightest objects since the observations of each pointing were taken in a series of scans with varying elevations.

\begin{figure*}
  \includegraphics[width=17.5cm]{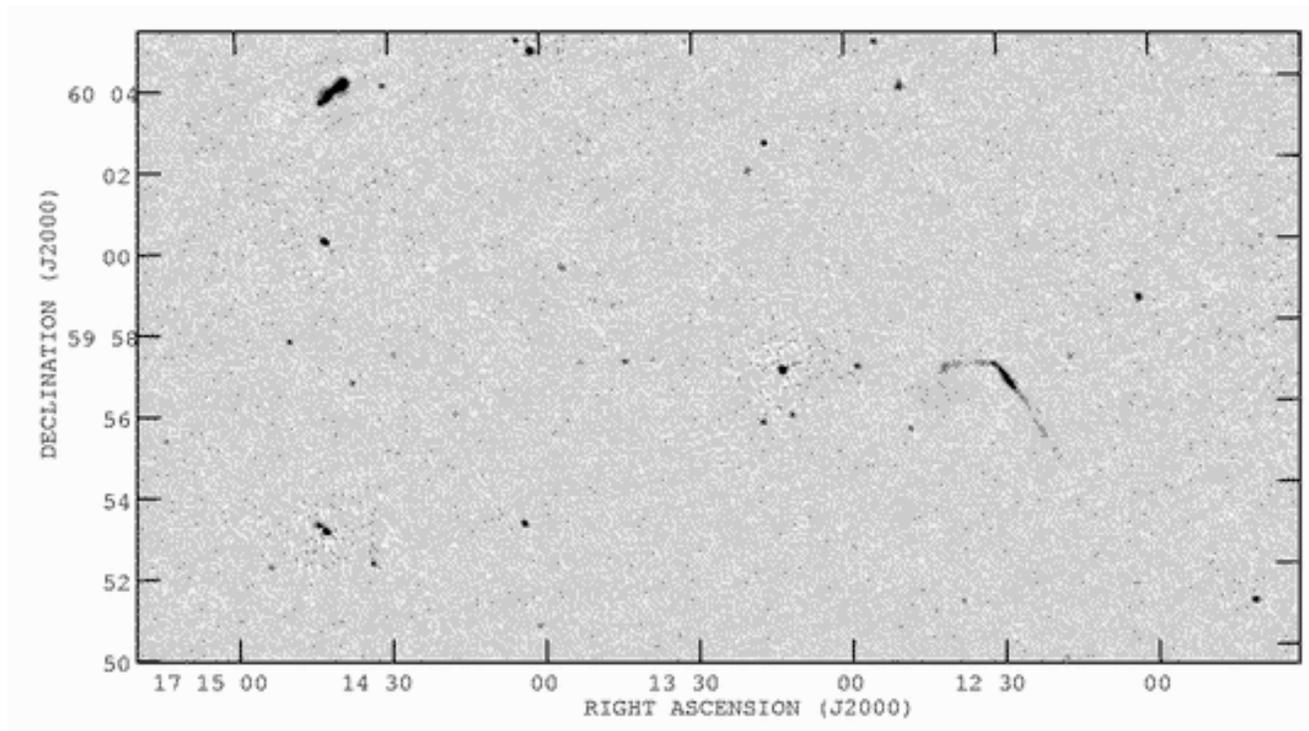}
  \caption{A sample section of the GMRT image.  The greyscale ranges between $-0.2$ and 1~mJy~beam$^{-1}$, and the resolution of the image is $5.8 \times 4.7$ arcsec$^2$, PA $+60^\circ$.  Most objects are unresolved, though one clearly extended object is visible near the middle.}
  \label{fig:GMRTarea}
\end{figure*}

\begin{figure*}
  \includegraphics[width=17.5cm]{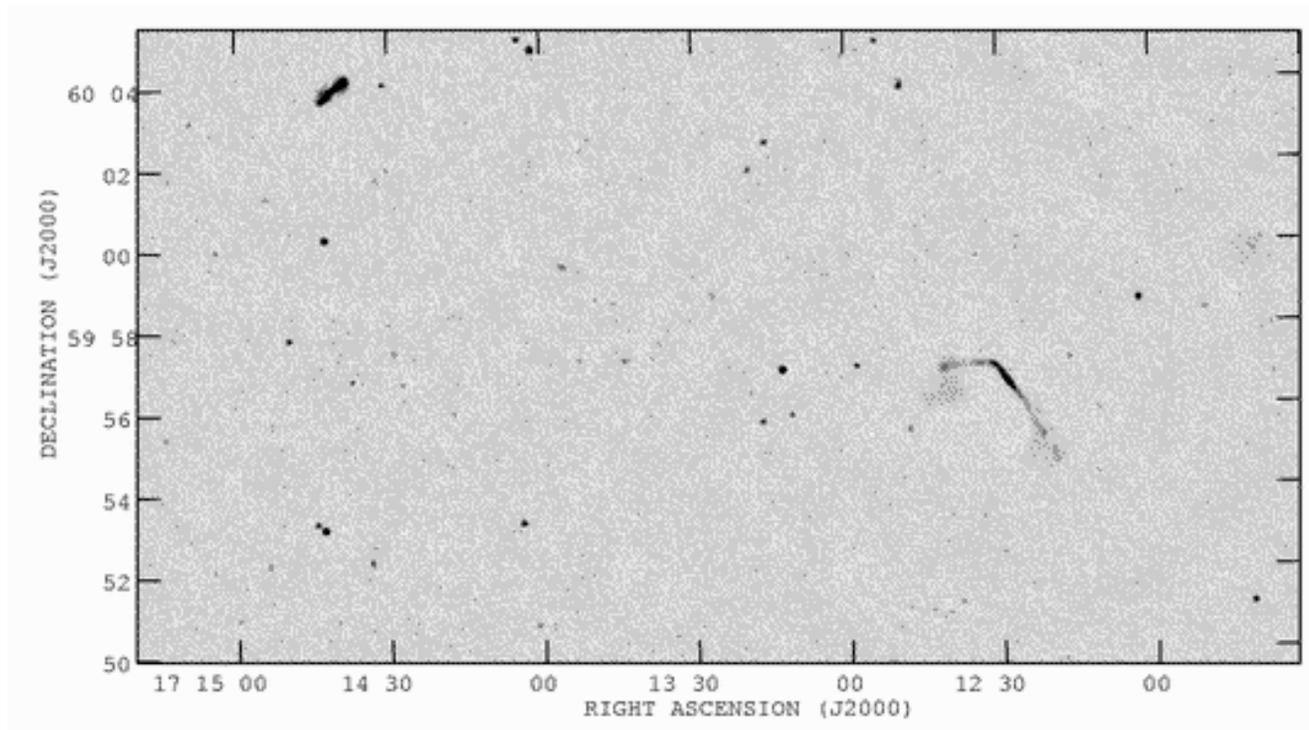}
  \caption{The existing 1.4-GHz VLA map for the same region \citep{Condon03}. The greyscale ranges between $-0.1$ and 0.5~mJy~beam$^{-1}$, equivalent to the GMRT image for objects with a spectral index of 0.8.  The image resolution is $5$~arcsec$^2$.}
  \label{fig:VLAarea}
\end{figure*}

A small portion of the mosaic is shown in Fig.~\ref{fig:GMRTarea}, demonstrating the quality of the image away from the bright sources, with the VLA map of the same area shown in Fig.~\ref{fig:VLAarea} for comparison.  The greyscale has been set on Figs~\ref{fig:GMRTarea} and \ref{fig:VLAarea} so that an object with a spectral index of 0.8 will appear equally bright in both images.  Most sources are unresolved in the 610-MHz image, although there are some objects present with extended structures -- we present a sample of these in Fig.~\ref{fig:images}.

\begin{figure*}
  \vskip1cm
  \centerline{\subfigure[FLSGMRT~J171247.5$+$591428.3]{
                \includegraphics[width=5.5cm]{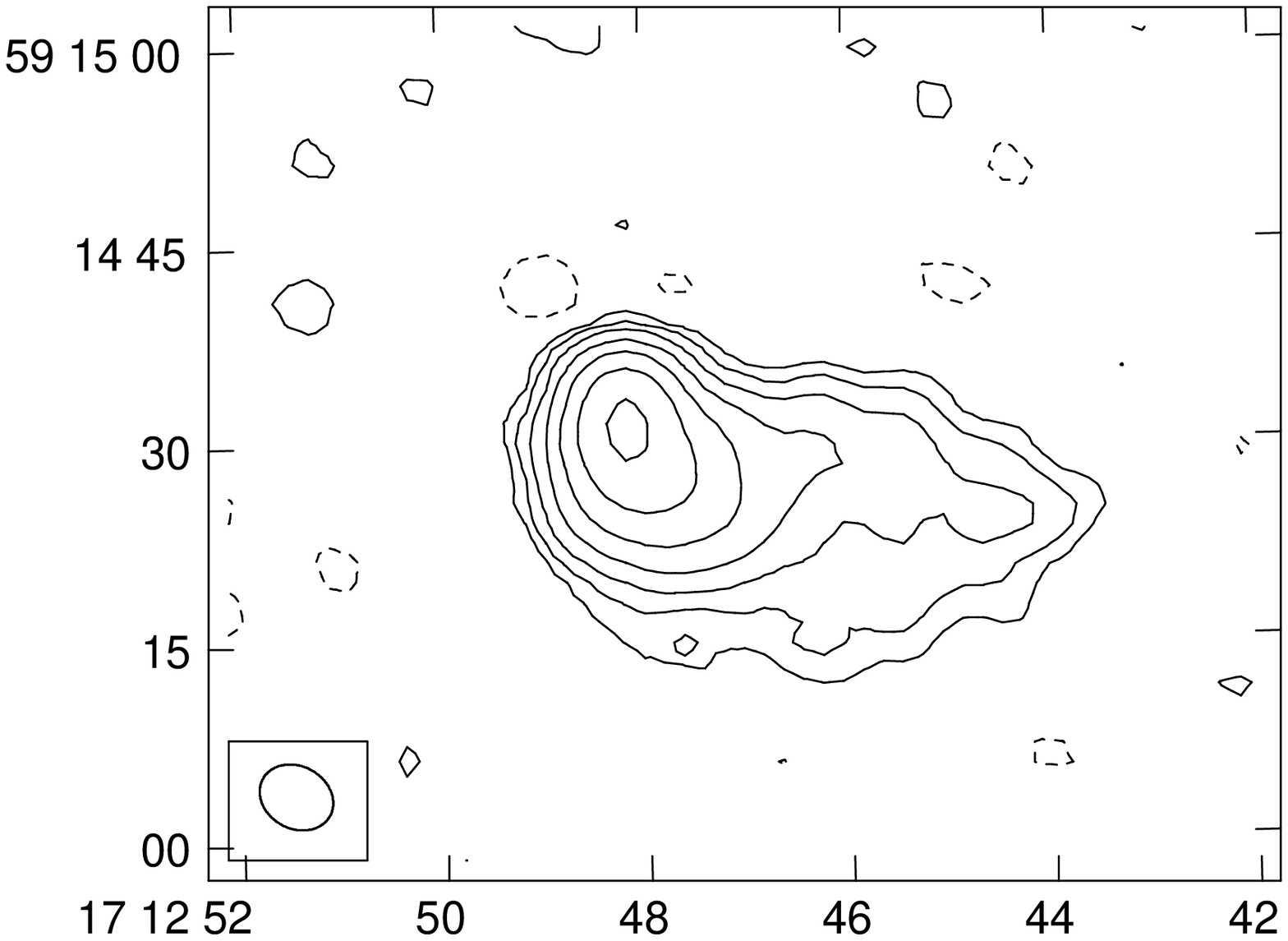}}
              \subfigure[FLSGMRT~J171440.2$+$600401]{
                \includegraphics[width=5.5cm]{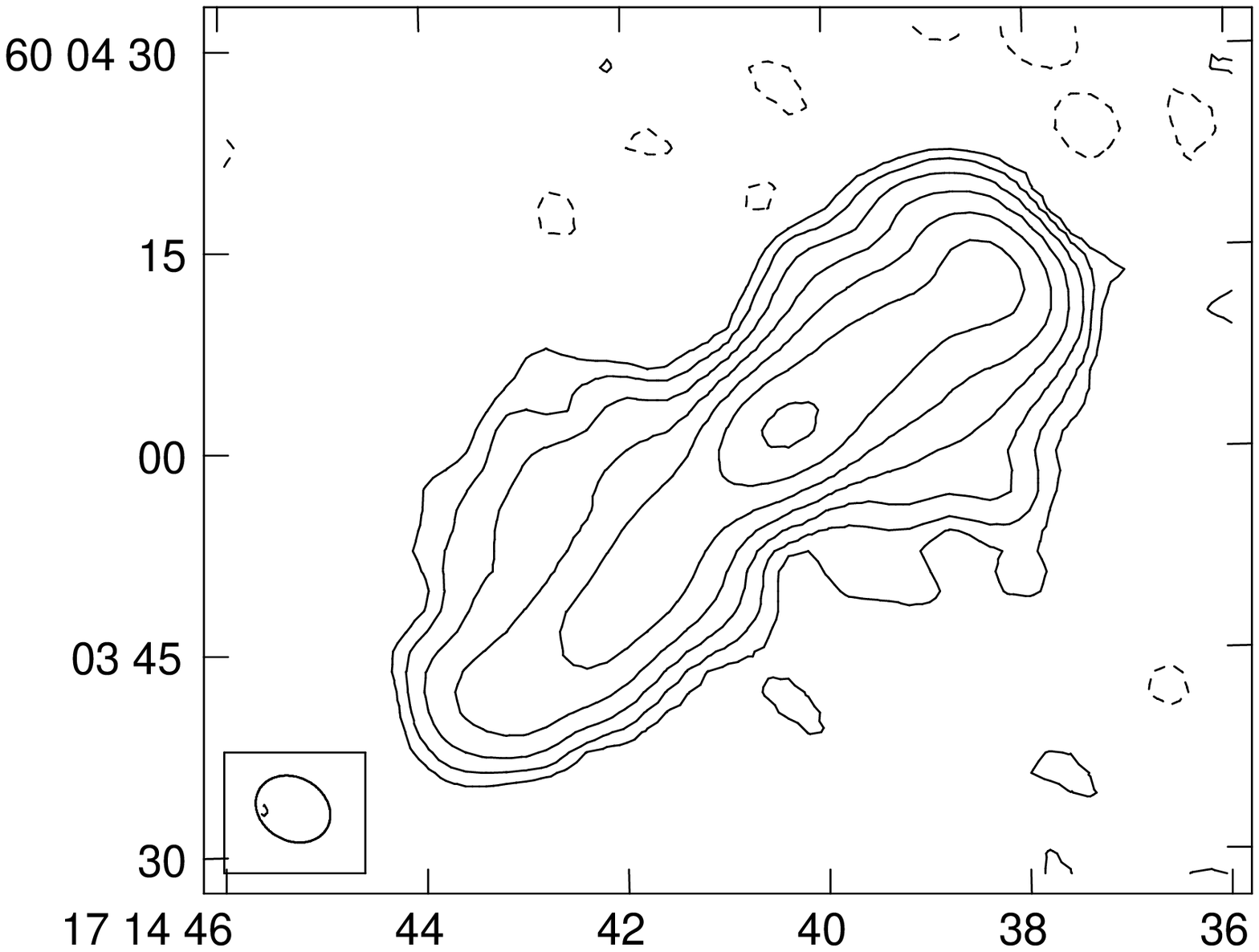}}
              \subfigure[FLSGMRT~J171437.0$+$594711 and J171435.1$+$594728]{
                \includegraphics[width=5.5cm]{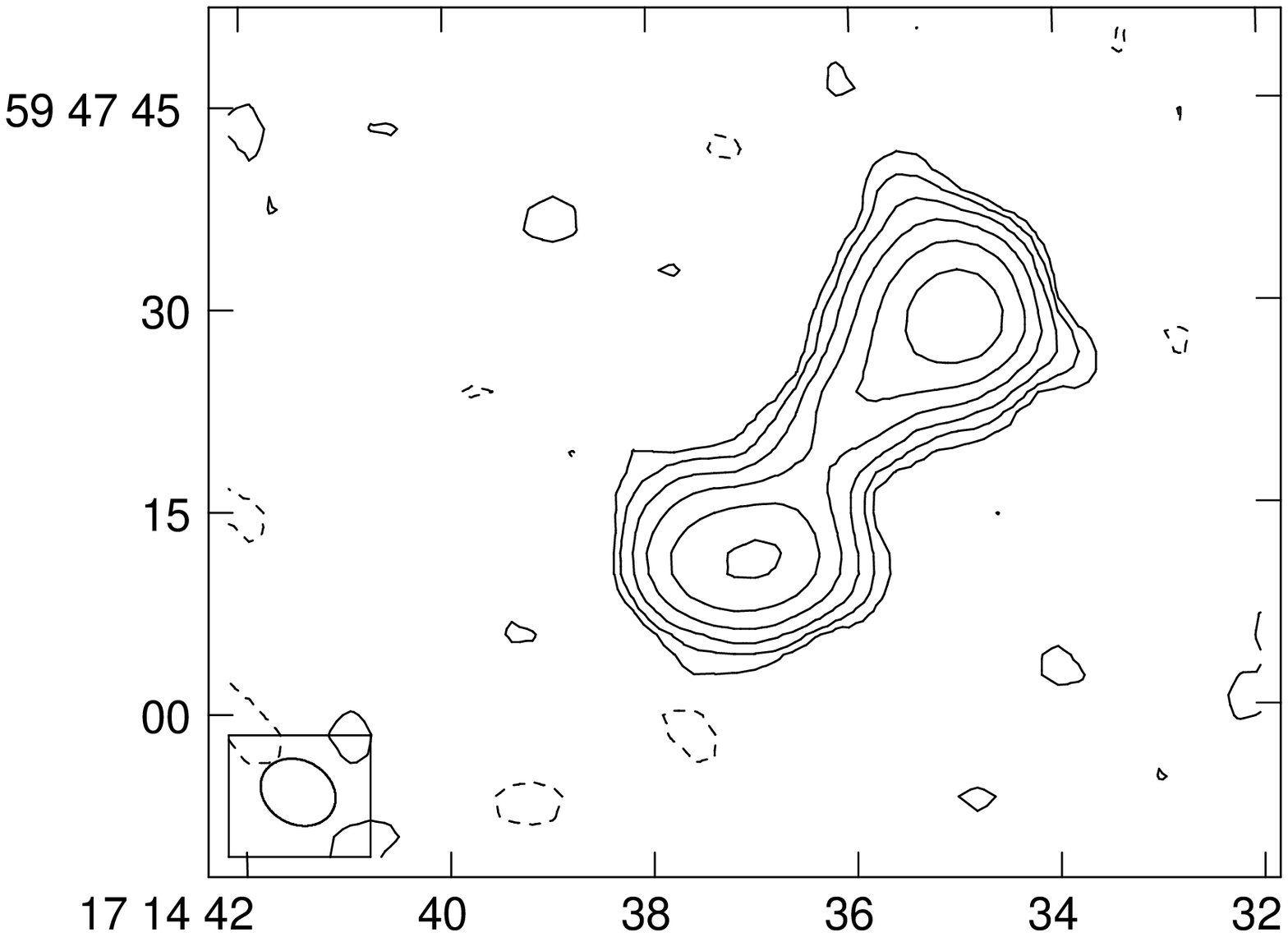}}
             }
  \centerline{\subfigure[FLSGMRT~J172340.4$+$593721 and J172336.4$+$593715]{
                \includegraphics[width=8cm]{IMG04.PS}}
              \hskip1cm
              \subfigure[FLSGMRT~J171735.9$+$600752 and J171731.4$+$600746]{
                \includegraphics[width=8cm]{IMG05.PS}}
             }
  \centerline{\subfigure[FLSGMRT~J171344.4$+$584533 and J171341.9$+$584540]{
                \includegraphics[width=5.3cm]{IMG06.PS}}
              \hskip0.5cm
              \subfigure[FLSGMRT~J171954.3$+$585234 and J171953.2$+$585222]{
                \includegraphics[width=5.3cm]{IMG07.PS}}
              \hskip0.5cm
              \subfigure[FLSGMRT~J171509.0$+$584526]{
                \includegraphics[width=5.3cm]{IMG08.PS}}
             }
  \centerline{\subfigure[FLSGMRT~J171413.8$+$594800 and J171414.6$+$594750]{
                \includegraphics[width=5.3cm]{IMG09.PS}}
              \hskip0.5cm
              \subfigure[FLSGMRT~J172439.4$+$592044, J172435.7$+$592053 and J172437.3$+$592017]{
                \includegraphics[width=5.3cm]{IMG10.PS}}
              \hskip0.5cm
              \subfigure[FLSGMRT~J172115.4$+$600432 and J172116.7$+$600406]{
                \includegraphics[width=5.3cm]{IMG11.PS}}
             }
  \caption{A selection of extended objects -- contours are plotted at $\pm$100 $\mu$Jy beam$^{-1}$ $\times$ 1, 2, 4, 8, $\ldots$  Negative contours are represented by dashed lines.  The resolution of the beam is shown in the bottom left of each image, and the designations of each source in the GMRT 610-MHz catalogue are given below.  All sources also appear extended in the VLA 1.4-GHz catalogue, and images are available in \citet{Condon03}.  }
  \label{fig:images}
\end{figure*}

\begin{figure*}
  \setcounter{subfigure}{11}
  \centerline{\subfigure[The extended source has been separated into FLSGMRT~J171240.3$+$595659, J171229.1$+$595639, J171223.0$+$595551 and J171222.2$+$595524.  The two compact sources are FLSGMRT~J171248.1$+$595528 and J171216.4$+$595711.]{
                \includegraphics[width=15cm]{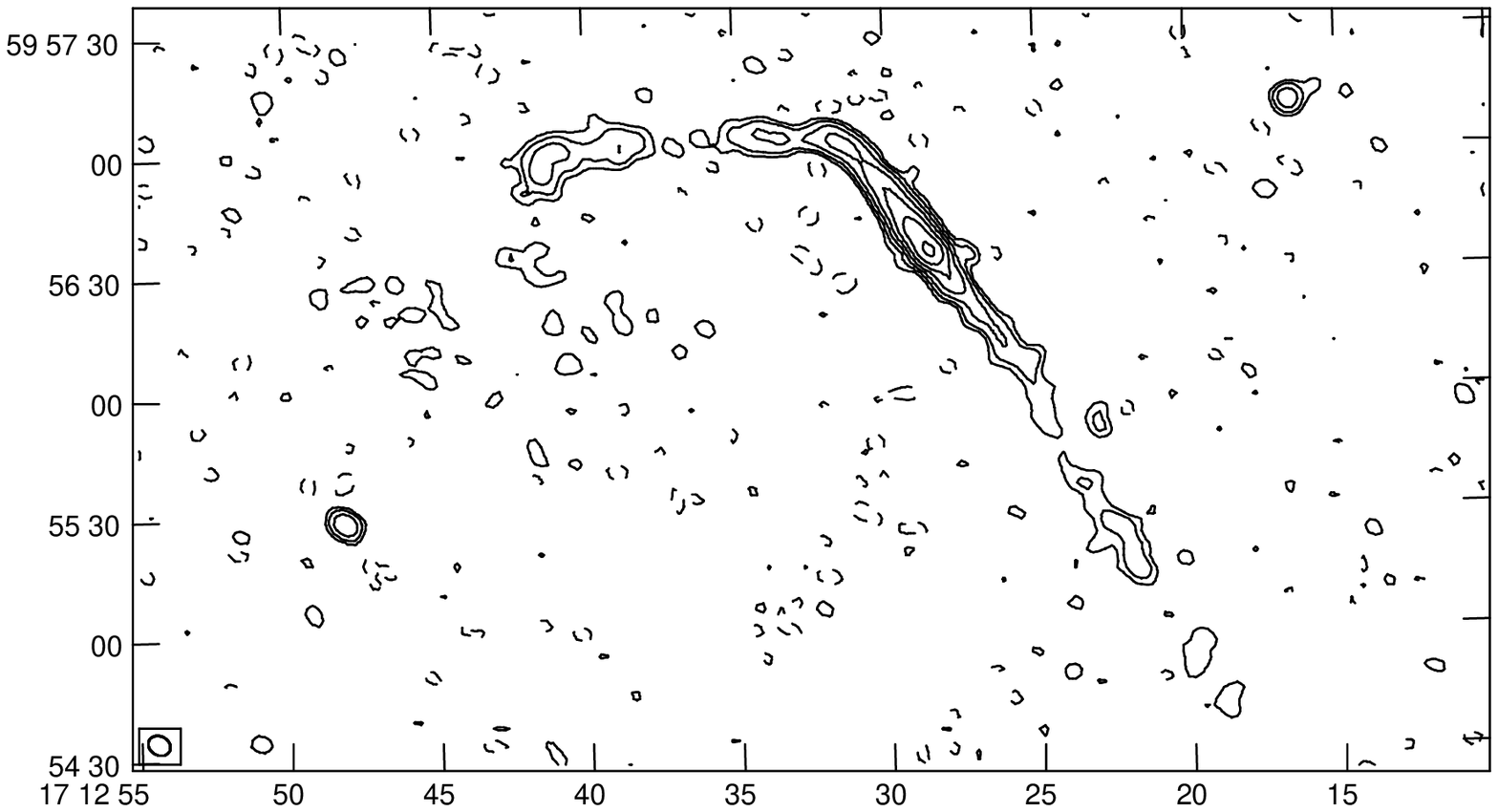}}
             }
  \centerline{\subfigure[FLSGMRT~J171147.7$+$600836, J171145.4$+$600808 and FLSGMRT~J171140.8$+$600731]{
                \includegraphics[width=7cm]{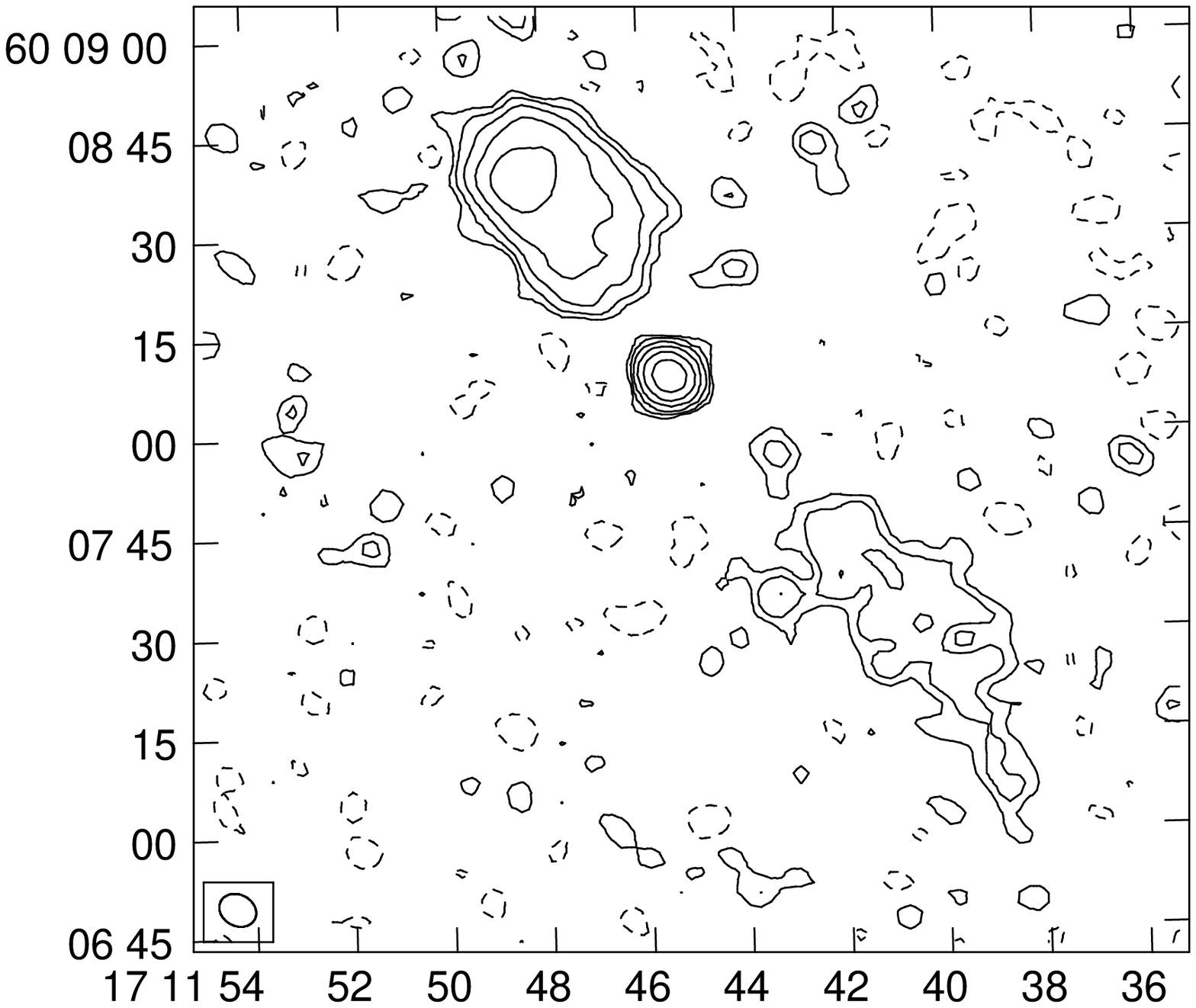}}
              \hskip1cm
              \subfigure[FLSGMRT~J171841.3$+$603630]{
                \includegraphics[width=7cm]{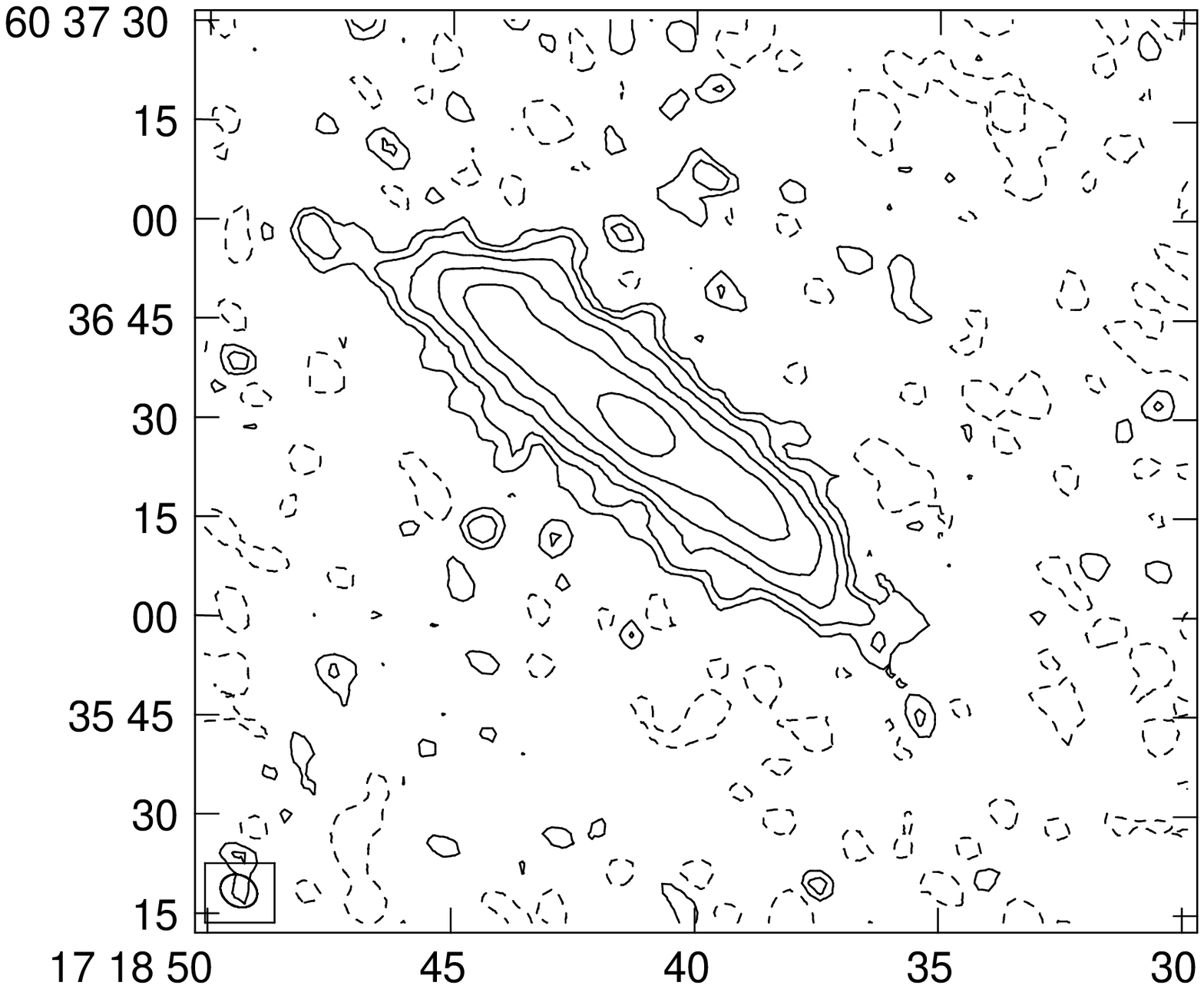}}
            }
  \centerline{\subfigure[The extended object has been separated into FLSGMRT~J171647.4$+$591458, J171639.4$+$151513, J171631.1$+$591544 and J171628.4$+$591527.  The double source components are FLSGMRT~J171622.4$+$591536 and J171623.1$+$591523.]{
                \includegraphics[width=15cm]{IMG15.PS}}
            }
  \contcaption{}
\end{figure*}

\begin{figure}
  \includegraphics[height=8cm,angle=-90]{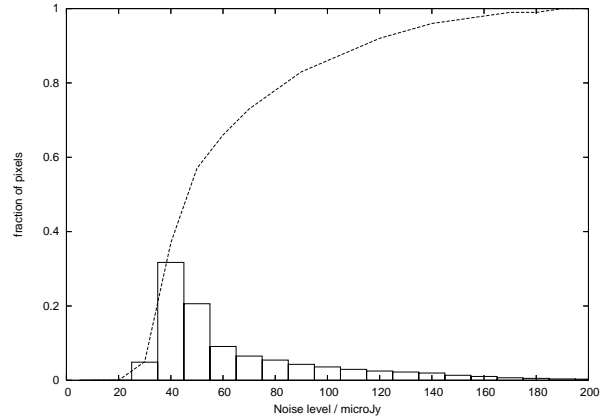}
  \caption{Pixel noise distibution, from the Source Extractor r.m.s.\ map.  The histogram shows the fraction of pixels at that noise, while the dotted line shows the cumulative fraction of pixels with noise below a particular value.}
  \label{fig:Ns}
\end{figure}

\begin{figure}
  \includegraphics[height=8cm,angle=-90]{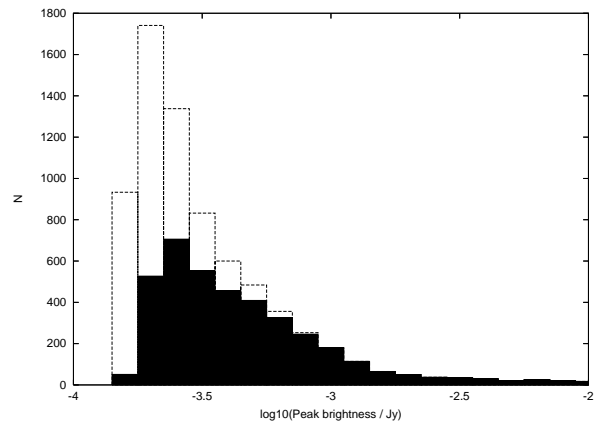}
  \caption{Source number counts from the 610-MHz Spitzer extragalactic First Look Survey field.  The solid boxes represent uncorrected counts, while the open boxes represent the counts after correction for the survey sensitivity level for each solid angle.}
  \label{fig:Peak1}
\end{figure}

A catalogue of 3944 sources was created using Source Extractor \citep{Bertin96} with peak brightness greater than $\sim5\sigma$ -- see Appendix~B for further details.  Source Extractor has a significant advantage over AIPS tasks such as SAD when creating a source list, in that it is capable of calculating the local background and noise level on the image.  Phase errors near the bright sources lead to an increase in noise, but by using a box of $16 \times 16$ pixels to estimate the local noise, the number of spurious detections in the final catalogue was reduced considerably -- this means that the fit is less deep near the brightest sources.

Table $\ref{tab:catalogue}$ presents a sample of 60 entries in the catalogue, which is sorted by right ascension.  The full table, and radio image of the Spitzer extragalactic First Look Survey field will be available via {\tt http://www.mrao.cam.ac.uk/surveys/}.  Column 1 gives the IAU designation of the source, in the form FLSGMRT~Jhhmmss.s$+$ddmmss, where J represents J2000.0 coordinates, hhmmss.s represents right ascension in hours, minutes and truncated tenths of seconds, and ddmmss represents the declination in degrees, arcminutes and truncated arcseconds.  Columns 2 and 3 give the right ascension and declination of the source, calculated by first moments of the relevant pixel flux densities to give a centroid position.  Column 4 gives the brightness of the peak pixel in each source, in mJy~beam$^{-1}$, and column 5 gives the local r.m.s.\ noise in $\mu$Jy~beam$^{-1}$.  Column 6 gives the integrated flux density in mJy, calculated from the mosaic and applying the flux density correction factor described in Appendix~B.  Column 7 gives the error in integrated flux density, calculated from the local noise level and source size, and taking into account the correction factor.  Columns 8 and 9 give the $X$, $Y$ pixel coordinates from the mosaic image of the source centroid.  Column 10 is the Source Extractor deblended object flag -- 0 for most objects, but 1 when a source has been split up into two or more components.  For deblended objects it is necessary to examine the image in order to distinguish between the case where two astronomically distinct objects have been split up, and when one extended object has been represented by more than one entry.  There are 211 deblended sources in our catalogue.

The non-uniform noise characteristics of our survey make it important to quantify the area that has been surveyed with each noise level.  Figure \ref{fig:Ns} shows the fraction of pixels with a particular noise level (taken from the Source Extractor r.m.s.\ map), and the cumulative fraction of pixels at that noise level.  The peak brightness distribution has been plotted in Fig.~\ref{fig:Peak1}, along with a distribution that has been corrected for the varying amount of solid angle being surveyed to each sensitivity level.  Previous studies have shown \citep{Hopkins02} that Source Extractor has a false detection rate of below 5\%, and detects above 90\% of sources with peak brightness close to the detection threshold.

\begin{table*}
  \caption{A sample of 60 entries from the 610-MHz extragalactic First Look Survey catalogue, sorted by right ascension.}
  \label{tab:catalogue}
  \begin{center}
  \begin{tabular}{cccccccccc} \hline \hline
Name & RA & Dec & Peak & Local Noise & Int. Flux Density & Error & X & Y & Flags\\
 & J2000.0 & J2000.0 & mJy~beam$^{-1}$ & $\mu$Jy beam$^{-1}$ & mJy & mJy & & & \\
(1)  & (2)             & (3)         & (4)       & (5)                & (6)      & (7)        & (8) & (9) & (10) \\
\hline
FLSGMRT J171751.2$+$584823 & 17:17:51.24 & $+$58:48:23.2 & 0.210 & 35 & 0.330 & 0.070 & 3486 & 1752 & 0\\
FLSGMRT J171751.3$+$594336 & 17:17:51.31 & $+$59:43:36.2 & 0.190 & 34 & 0.150 & 0.050 & 3486 & 3961 & 0\\
FLSGMRT J171751.3$+$591436 & 17:17:51.31 & $+$59:14:36.5 & 0.210 & 39 & 0.140 & 0.050 & 3486 & 2801 & 0\\
FLSGMRT J171751.3$+$590351 & 17:17:51.34 & $+$59:03:51.1 & 0.380 & 41 & 0.490 & 0.080 & 3486 & 2371 & 0\\
FLSGMRT J171751.5$+$590017 & 17:17:51.57 & $+$59:00:17.8 & 0.290 & 42 & 0.350 & 0.070 & 3484 & 2228 & 0\\
FLSGMRT J171751.6$+$590931 & 17:17:51.69 & $+$59:09:31.2 & 0.310 & 42 & 0.350 & 0.070 & 3484 & 2597 & 0\\
FLSGMRT J171751.9$+$591446 & 17:17:51.96 & $+$59:14:46.4 & 0.400 & 40 & 0.470 & 0.070 & 3483 & 2807 & 0\\
FLSGMRT J171752.0$+$603620 & 17:17:52.00 & $+$60:36:20.1 & 0.310 & 57 & 0.350 & 0.100 & 3483 & 6070 & 0\\
FLSGMRT J171752.4$+$594157 & 17:17:52.44 & $+$59:41:57.9 & 0.280 & 37 & 0.310 & 0.070 & 3480 & 3895 & 0\\
FLSGMRT J171752.5$+$600451 & 17:17:52.59 & $+$60:04:51.1 & 0.230 & 38 & 0.240 & 0.060 & 3480 & 4811 & 0\\
FLSGMRT J171752.9$+$591241 & 17:17:52.95 & $+$59:12:41.7 & 0.440 & 43 & 0.480 & 0.070 & 3478 & 2724 & 0\\
FLSGMRT J171753.0$+$584608 & 17:17:53.01 & $+$58:46:08.4 & 0.320 & 38 & 0.610 & 0.080 & 3477 & 1662 & 0\\
FLSGMRT J171753.1$+$601508 & 17:17:53.11 & $+$60:15:08.1 & 0.200 & 33 & 0.260 & 0.060 & 3477 & 5222 & 0\\
FLSGMRT J171753.1$+$592330 & 17:17:53.14 & $+$59:23:30.3 & 0.180 & 32 & 0.300 & 0.070 & 3477 & 3157 & 0\\
FLSGMRT J171753.3$+$585502 & 17:17:53.37 & $+$58:55:02.3 & 0.290 & 35 & 0.370 & 0.070 & 3475 & 2018 & 0\\
FLSGMRT J171753.5$+$594912 & 17:17:53.54 & $+$59:49:12.2 & 0.350 & 40 & 0.410 & 0.070 & 3475 & 4185 & 0\\
FLSGMRT J171753.7$+$594428 & 17:17:53.74 & $+$59:44:28.2 & 0.230 & 41 & 0.180 & 0.060 & 3474 & 3995 & 0\\
FLSGMRT J171753.7$+$592117 & 17:17:53.75 & $+$59:21:18.0 & 0.180 & 34 & 0.190 & 0.060 & 3474 & 3069 & 0\\
FLSGMRT J171753.9$+$594300 & 17:17:53.98 & $+$59:43:00.5 & 0.520 & 35 & 0.570 & 0.060 & 3473 & 3937 & 1\\
FLSGMRT J171754.7$+$593533 & 17:17:54.73 & $+$59:35:33.4 & 0.200 & 35 & 0.170 & 0.050 & 3469 & 3639 & 0\\
FLSGMRT J171754.7$+$600914 & 17:17:54.73 & $+$60:09:14.8 & 0.670 & 32 & 0.890 & 0.070 & 3469 & 4986 & 0\\
FLSGMRT J171754.8$+$601939 & 17:17:54.84 & $+$60:19:39.4 & 0.160 & 29 & 0.180 & 0.050 & 3469 & 5403 & 0\\
FLSGMRT J171755.0$+$585933 & 17:17:55.03 & $+$58:59:33.7 & 0.290 & 44 & 0.560 & 0.110 & 3467 & 2199 & 0\\
FLSGMRT J171755.0$+$594300 & 17:17:55.05 & $+$59:43:00.5 & 0.190 & 35 & 0.130 & 0.040 & 3467 & 3937 & 1\\
FLSGMRT J171755.2$+$604141 & 17:17:55.29 & $+$60:41:41.4 & 0.420 & 76 & 0.270 & 0.100 & 3467 & 6284 & 0\\
FLSGMRT J171755.2$+$585512 & 17:17:55.29 & $+$58:55:12.8 & 0.360 & 33 & 0.410 & 0.060 & 3465 & 2025 & 0\\
FLSGMRT J171755.3$+$592846 & 17:17:55.32 & $+$59:28:46.7 & 0.280 & 30 & 0.440 & 0.060 & 3466 & 3368 & 0\\
FLSGMRT J171755.6$+$594815 & 17:17:55.62 & $+$59:48:15.4 & 0.220 & 38 & 0.180 & 0.050 & 3464 & 4147 & 0\\
FLSGMRT J171755.6$+$593421 & 17:17:55.63 & $+$59:34:21.6 & 0.180 & 29 & 0.120 & 0.040 & 3464 & 3591 & 0\\
FLSGMRT J171755.8$+$600519 & 17:17:55.83 & $+$60:05:19.3 & 0.190 & 34 & 0.260 & 0.070 & 3464 & 4829 & 1\\
FLSGMRT J171756.0$+$594431 & 17:17:56.08 & $+$59:44:31.9 & 0.310 & 40 & 0.460 & 0.080 & 3462 & 3998 & 0\\
FLSGMRT J171756.4$+$592839 & 17:17:56.40 & $+$59:28:39.6 & 0.170 & 31 & 0.170 & 0.050 & 3460 & 3363 & 0\\
FLSGMRT J171756.4$+$591744 & 17:17:56.41 & $+$59:17:44.2 & 0.310 & 41 & 0.400 & 0.070 & 3460 & 2926 & 0\\
FLSGMRT J171756.4$+$590005 & 17:17:56.49 & $+$59:00:05.4 & 0.450 & 43 & 0.560 & 0.080 & 3459 & 2220 & 0\\
FLSGMRT J171756.5$+$584809 & 17:17:56.51 & $+$58:48:09.1 & 0.230 & 39 & 0.300 & 0.070 & 3459 & 1743 & 0\\
FLSGMRT J171756.8$+$593631 & 17:17:56.85 & $+$59:36:31.1 & 0.470 & 32 & 0.580 & 0.060 & 3458 & 3677 & 0\\
FLSGMRT J171756.9$+$601401 & 17:17:56.90 & $+$60:14:01.1 & 0.190 & 33 & 0.210 & 0.050 & 3459 & 5177 & 0\\
FLSGMRT J171757.5$+$602115 & 17:17:57.54 & $+$60:21:15.7 & 0.200 & 32 & 0.290 & 0.060 & 3456 & 5467 & 0\\
FLSGMRT J171757.6$+$594247 & 17:17:57.61 & $+$59:42:47.4 & 0.310 & 38 & 0.310 & 0.060 & 3454 & 3928 & 0\\
FLSGMRT J171757.6$+$584555 & 17:17:57.65 & $+$58:45:56.0 & 0.670 & 37 & 0.930 & 0.080 & 3453 & 1654 & 0\\
FLSGMRT J171757.7$+$593136 & 17:17:57.80 & $+$59:31:36.3 & 0.380 & 33 & 0.390 & 0.050 & 3453 & 3481 & 0\\
FLSGMRT J171757.7$+$591527 & 17:17:57.80 & $+$59:15:27.4 & 0.200 & 38 & 0.140 & 0.050 & 3453 & 2835 & 0\\
FLSGMRT J171757.9$+$585604 & 17:17:57.93 & $+$58:56:04.3 & 0.240 & 39 & 0.260 & 0.070 & 3452 & 2059 & 0\\
FLSGMRT J171758.1$+$604204 & 17:17:58.18 & $+$60:42:04.6 & 0.450 & 77 & 0.420 & 0.110 & 3453 & 6299 & 0\\
FLSGMRT J171758.2$+$585811 & 17:17:58.26 & $+$58:58:11.9 & 0.230 & 39 & 0.360 & 0.080 & 3450 & 2145 & 0\\
FLSGMRT J171758.3$+$602850 & 17:17:58.37 & $+$60:28:50.8 & 0.300 & 48 & 0.400 & 0.090 & 3452 & 5770 & 0\\
FLSGMRT J171758.7$+$590407 & 17:17:58.76 & $+$59:04:07.4 & 0.260 & 45 & 0.230 & 0.060 & 3448 & 2382 & 0\\
FLSGMRT J171758.9$+$602307 & 17:17:58.93 & $+$60:23:07.7 & 0.260 & 36 & 0.410 & 0.070 & 3449 & 5542 & 0\\
FLSGMRT J171758.9$+$593454 & 17:17:58.96 & $+$59:34:54.5 & 0.150 & 29 & 0.130 & 0.040 & 3447 & 3613 & 0\\
FLSGMRT J171759.0$+$592438 & 17:17:59.08 & $+$59:24:38.9 & 1.910 & 33 & 2.280 & 0.070 & 3446 & 3202 & 0\\
FLSGMRT J171759.1$+$584435 & 17:17:59.11 & $+$58:44:35.8 & 0.440 & 39 & 0.680 & 0.080 & 3445 & 1600 & 0\\
FLSGMRT J171759.3$+$594325 & 17:17:59.38 & $+$59:43:25.7 & 0.210 & 39 & 0.460 & 0.100 & 3445 & 3954 & 0\\
FLSGMRT J171759.4$+$590132 & 17:17:59.40 & $+$59:01:32.1 & 0.260 & 38 & 0.230 & 0.050 & 3444 & 2278 & 0\\
FLSGMRT J171759.3$+$595246 & 17:17:59.40 & $+$59:52:46.7 & 0.980 & 42 & 1.250 & 0.090 & 3446 & 4328 & 0\\
FLSGMRT J171759.4$+$590256 & 17:17:59.43 & $+$59:02:56.4 & 0.210 & 40 & 0.290 & 0.070 & 3444 & 2334 & 0\\
FLSGMRT J171759.8$+$594210 & 17:17:59.84 & $+$59:42:10.8 & 0.220 & 37 & 0.230 & 0.060 & 3443 & 3904 & 0\\
FLSGMRT J171800.0$+$584604 & 17:18:00.04 & $+$58:46:04.5 & 0.250 & 40 & 0.260 & 0.060 & 3440 & 1660 & 0\\
FLSGMRT J171800.1$+$594925 & 17:18:00.16 & $+$59:49:25.1 & 0.250 & 42 & 0.360 & 0.080 & 3442 & 4193 & 0\\
FLSGMRT J171800.3$+$590207 & 17:18:00.31 & $+$59:02:07.8 & 1.220 & 40 & 1.510 & 0.090 & 3440 & 2302 & 0\\
FLSGMRT J171800.3$+$585142 & 17:18:00.35 & $+$58:51:42.7 & 0.190 & 37 & 0.330 & 0.080 & 3439 & 1885 & 0\\
\hline
  \end{tabular}
  \end{center}
\end{table*}

%------------------------------------------------------------------------------%
\section{Comparison with the VLA survey}
The VLA survey of the xFLS region \citep{Condon03} has a uniform noise level of 23~$\mu$Jy.  This is equivalent to a noise of $\sim45~\mu$Jy at 610~MHz, assuming a spectral index of 0.8.  Our observations have significantly lower noise levels than this in the centre of our mosaic, with the noise in the overlap regions between two pointings being approximately 45~$\mu$Jy (see Fig.~\ref{fig:sevenpoints}).  Figure~\ref{fig:Ns} shows that the noise is below 45~$\mu$Jy for about half the area of our mosaic.

\begin{figure}
  \includegraphics[width=8cm]{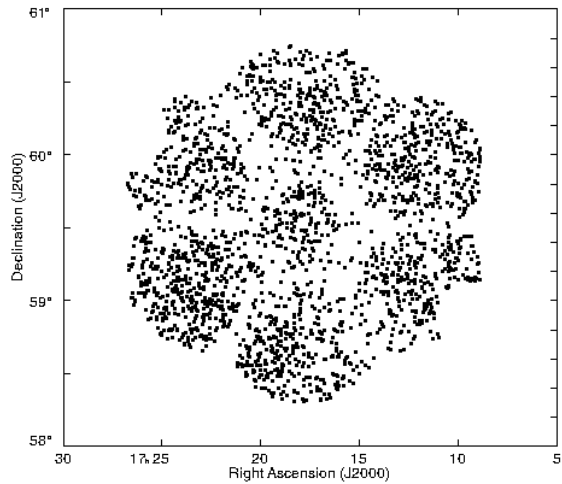}
  \caption{Source positions for objects detected at 610~MHz and undetected at 1.4~GHz.  Most sources are located near the centre of one of our pointings, where the GMRT sensitivity is greatest.}
  \label{fig:GMRTrejects}
\end{figure}

In order to make a quantitative comparison between the VLA xFLS survey and our GMRT survey, we have run Source Extractor on our mosaic and the 1.4-GHz image.  At the $\sim$5$\sigma$ level, there were 3826 sources detected at 610~MHz, and 3091 detected at 1.4~GHz, in the region covered by both surveys.  The source lists were matched using a pairing radius of $6''$, which is approximately the resolution of the two catalogues, and 1580 unique matches were found.  Figure~\ref{fig:GMRTrejects} shows the source distribution of objects found by the GMRT but not by the VLA -- the majority of the unmatched sources are found in regions where our observations are deeper than the VLA image.

It is likely that some of the sources that are not detected at 610~MHz but are detected at 1.4~GHz are flat spectrum objects, with spectral index $\leq 0.5$.  The VLA noise of 23~$\mu$Jy would be equivalent to a 610-MHz noise level of $\sim35~\mu$Jy for $\alpha = 0.5$, so faint flat spectrum objects detectable throughout the VLA image would only be detectable in a small region of our mosaic.

\begin{figure}
  \includegraphics[height=8cm,width=8cm,angle=-90]{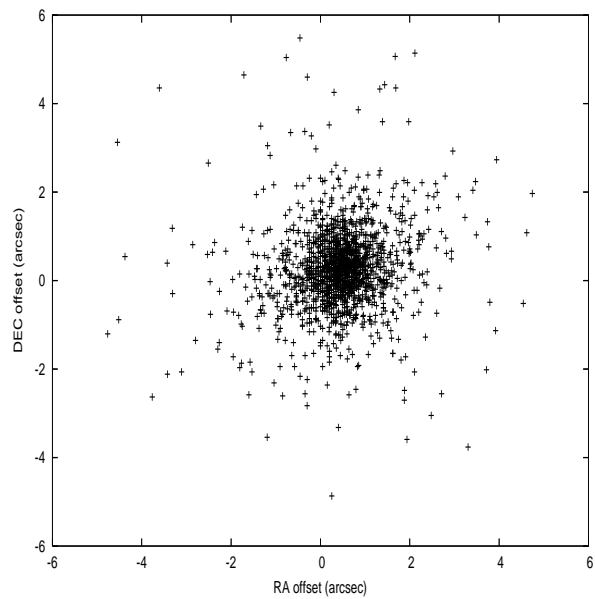}
  \caption{Source positions in the GMRT catalogue relative to the positions found in the VLA image, for unique matches within $6''$.}
  \label{fig:deltaRADEC}
\end{figure}

\begin{figure}
  \includegraphics[height=8cm,angle=-90]{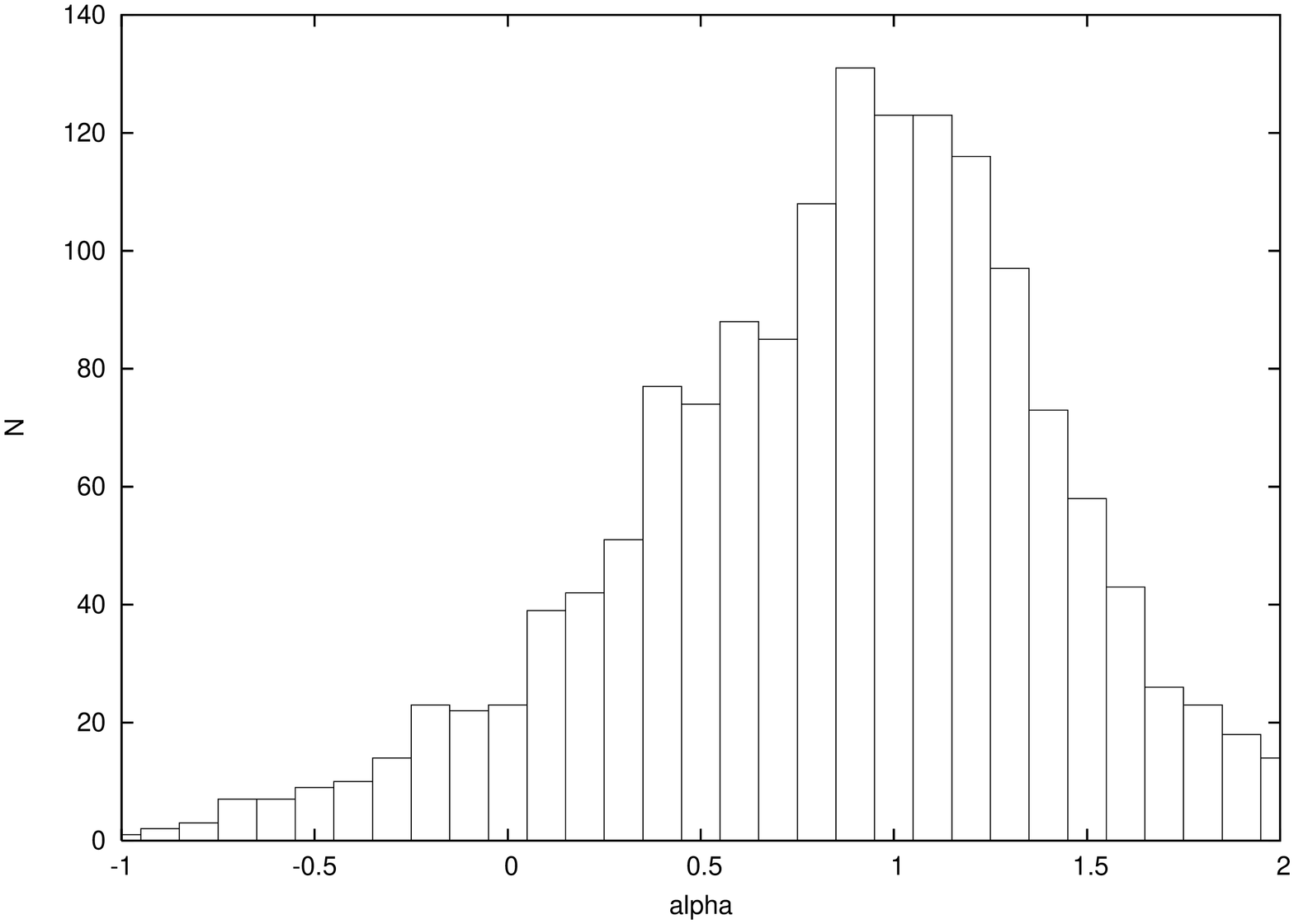}
  \caption{Radio spectral index between 610~MHz and 1.4~GHz}
  \label{fig:alpha}
\end{figure}

Figure~\ref{fig:deltaRADEC} shows the position offsets of the matched GMRT sources compared with their VLA counterparts.  The offsets have an approximately Gaussian distribution, with mean offset in right ascension of $0\farcs4$, standard deviation $0\farcs5$ and in declination $0\farcs2$ with a standard deviation of $0\farcs6$.  These offsets have not been applied to our catalogue, since it is uncertain as to which survey the errors come from.

The spectral index distribution of objects detected at both frequencies is shown in Fig.~\ref{fig:alpha}.  The integrated flux densities of sources have been used for the calculation, where the flux density of VLA sources was corrected using the same method as for the GMRT image (details in Appendix~B).

%------------------------------------------------------------------------------%
\section*{Acknowledgments}

We thank the staff of the GMRT who have made these observations possible.  TG thanks the UK PPARC for a Studentship.  The GMRT is operated by the National Centre for Radio Astrophysics of the Tata Institute of Fundamental Research, India.

%------------------------------------------------------------------------------%

%------------------------------------------------------------------------------%
\appendix

\section{Corrections to GMRT data}

\subsection{Corrections to the $uv$ data}

Two corrections were made to the GMRT $uv$ data before the final images were made.
\begin{enumerate}
  \item The $uv$ coverage was increased by combining the two sidebands into a single data file.  The frequency channels in the lower sideband (LSB) have a negative frequency increment, while the upper sideband (USB) has a positive increment -- this means that the two channels can not be combined directly using standard AIPS tasks.  We wrote an AIPS task called {\sc UVFLP} to re-order the LSB data, so that it was on the same frequency scale as the USB points, and so the sidebands could be combined before imaging.
  \item The coordinates of sources originally found in our GMRT images were seen to be slightly rotated near the edge of each pointing compared with their VLA positions.  This was due to incorrect time-stamps being used in the GMRT online software, leading to a slight error in the $uv$ data.  We wrote a customised AIPS task, {\sc UVFXT}, to increase the time-stamps by 7~s and correctly recompute the $uvw$ co-ordinates.  This was performed before the final images were created.
\end{enumerate}

\subsection{Primary beam correction}

During the preliminary analysis of the data, we compared the properties of several hundred relatively bright sources with peaks greater than 1~mJy~beam$^{-1}$, visible in the overlapping regions between two pointings.  This comparison revealed a systematic difference between the apparent brightness of sources in adjacent pointings; sources located to the north-west of a pointing were consistently brighter than the same source when viewed in the south-east of an adjacent pointing.  We were able to model this as the effective, average pointing centre of the telescope being offset by $\sim2.5$~arcmin in a north-west direction, compared with the nominal pointing centre.  The amount and direction of the offset was consistent between all pairs of pointings.  After applying the correction, the systematic effects were removed and the r.m.s.\ flux density errors were below 10\% near the edge of the primary beam, compared with nearer 20\% before the correction was applied.

It is thought that the primary beam offset depends on telescope elevation.  Our scans of each source were not taken symmetrically about source transit, and the detected offset is likely to be an average over the individual offsets of each scan.  Sources will have a slightly different position and flux density in each scan, due to this elevation-dependant error.  Self-calibration assumes that the source remains constant with time, and this is likely to be the reason why residual artefacts are seen near the brightest sources.

\begin{figure}
  \includegraphics[height=8cm,angle=-90]{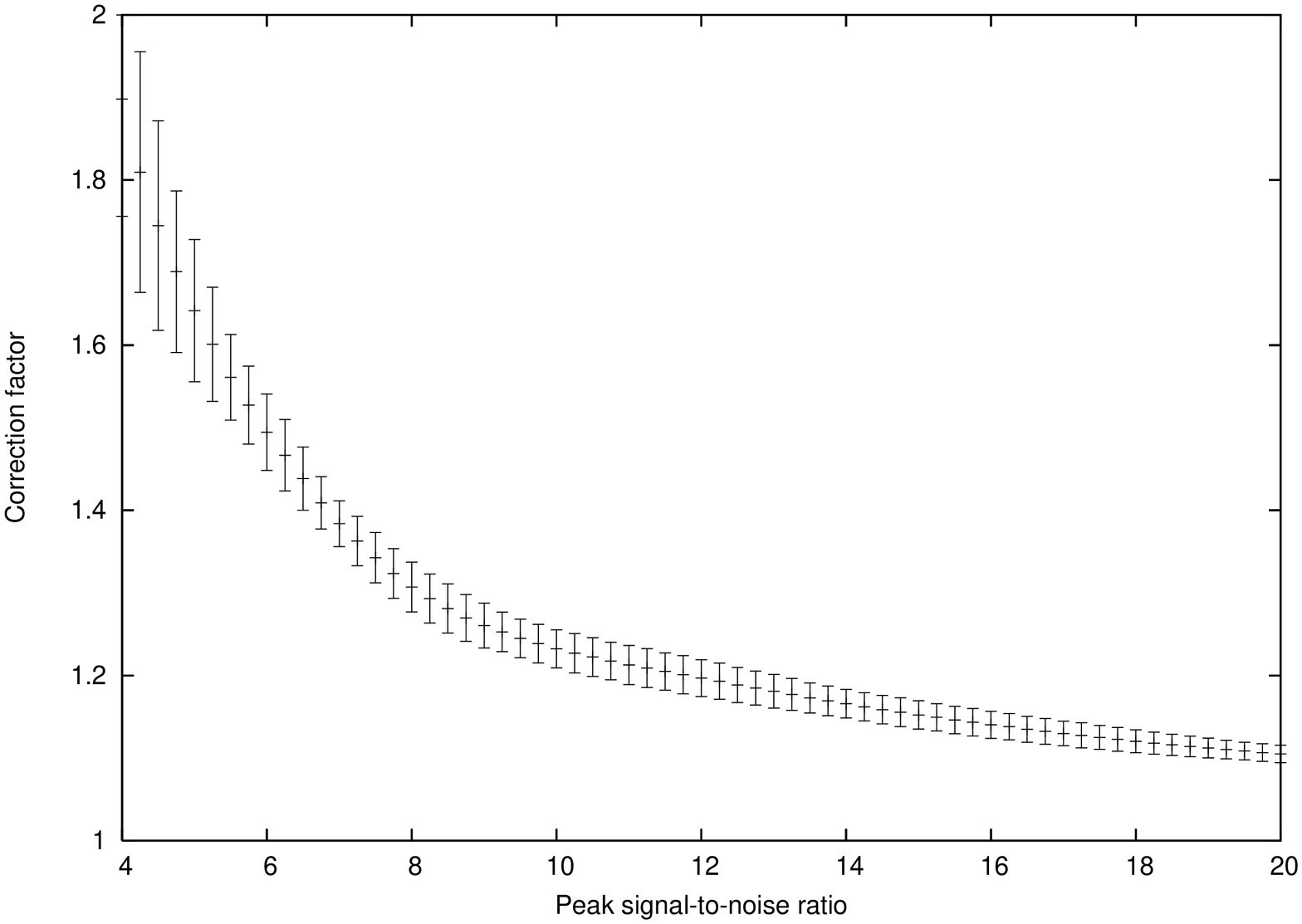}
  \caption{Flux density correction factor for point sources}\label{fig:corfact}
\end{figure}

\section{Source Extractor}

Source Extractor calculates the flux density of an object by summing all pixels greater than some user-defined threshold.  For an object to be included in our catalogue, we required it to have at least five connected pixels with brightness above 2$\sigma$, and a peak pixel brightness of greater than $5.25\sigma$.  Because of the oversampling of the beam, the peak of a source was taken to be the value of the brightest pixel in the island of flux.  The peak brightness requirement for our catalogue was set to the slightly more conservative $5.25\sigma$ rather than 5$\sigma$ due to the increased number of spurious sources near the edges of the image that were being detected at the lower cutoff.  3944 sources were identified and included in the catalogue.

The integrated flux density measured by Source Extractor only comes from pixels above 2$\sigma$.  This means that bright sources with a large signal-to-noise ratio (SNR) will have almost all their flux measured, whereas for faint sources an appreciable fraction of the flux is missed.  The required correction factor will depend only on the peak SNR, for unresolved sources.  We modelled a point source with a range of SNRs, and varied the exact location of the centre of the source across a pixel.  Figure~\ref{fig:corfact} shows the correction factor that has been applied to an unresolved source in order to obtain the true flux density.  In our catalogue we give only the corrected flux densities -- for extended objects that do not have the same shape as the beam, it is necessary to examine the image to obtain an accurate flux density.

%------------------------------------------------------------------------------%
\end{document}